\documentclass[multphys,vecphys]{svmult}

\usepackage{makeidx}         
\usepackage{graphicx}        
\usepackage{multicol}        
\usepackage[bottom]{footmisc}

\makeindex             

\begin{document}

\title*{High sensitivity low frequency radio observations of cD galaxies}
\author{S.~Giacintucci\inst{1,2,3}\and
T.~Venturi\inst{1}\and
S.~Bardelli\inst{3}\and
D.~Dallacasa\inst{1,2}\and
P.~Mazzotta\inst{4,5}\and
D.J.~Saikia\inst{6}}
\institute{INAF - IRA, via Gobetti 101, I-40129, Bologna, Italy \texttt{sgiaci$_{-}$s$@$ira.inaf.it}
\and Dip. Astronomia, Univ. Bologna, via Ranzani 1, I-40126, Bologna, Italy
\and INAF - OAB, via Ranzani 1, I-40126, Bologna, Italy
\and Univ. Roma Tor Vergata, via della Ricerca Scientifica 1, I-00133 Roma, Italy
\and Harvard-Smithsonian CfA, 60 Garden Street, Cambridge, MA 02138, USA
\and NCRA, Pune University campus, Ganeshkhind, Pune, 411 007, India}

%
%
\maketitle
 
\begin{abstract}
We present the GMRT 235 MHz images of three radio galaxies 
and 610 MHz images of two sources belonging to a complete 
sample of cD galaxies in rich and poor galaxy clusters.
The analysis of the spectral properties confirms the 
presence of aged radio emission in two of the presented 
sources.
\end{abstract}

\section{Introduction}
\label{sec:1}

cD galaxies are the most luminous and 
massive galaxies known. They are found at the centre of 
both rich and poor galaxy clusters, usually at the peak 
of the X--ray emission. Their radio emission shows a 
large variety of morphologies: compact, FRI and FRI/FRII 
morphology, wide--angle--tail (WAT), 
core--halo and more peculiar structures. 
The advent of the high resolution X--ray imaging with 
{\it Chandra} and XMM--Newton posed new questions related 
to the interaction between the central radio source and 
the intracluster medium (ICM). It is now clear that 
the radio lobes may displace the external medium and create 
depressions ({\it cavities}) in the ICM. These can be filled 
with radio emission ({\it radio--filled cavities}); 
alternatively a misplacement between radio lobes and 
cavities can be observed ({\it ghost cavities}). 
Both features may be present in the same cluster 
(e.g. Perseus; Fabian et al. \cite{fabian00}). In particular 
ghost cavities tend to be more external than the radio filled 
ones. This led to the suggestion that repeated outbursts of 
the central radio galaxy may play a role in the cavity formation 
(McNamara et al. \cite{mcnamara01}): ghost cavities would be 
the result of a previous AGN burst, and thus filled with aged 
radio plasma emitting at very low radio frequencies.
All recent investigations of the radio galaxy--ICM 
interaction in clusters mostly concentrated on single objects, 
based on the X--ray information. An unbiased comprehensive 
study of a statistical sample of clusters is still missing.

\subsection{cD sample and GMRT observations}
\label{sec:2}

In order to overcome possible biases induced by radio 
and/or X--ray selections, we defined a statistical sample 
of cDs starting from the optical information. 
We selected a sample of 132 cDs (109 in Abell clusters 
and 23 in poor clusters; for further details see Giacintucci 
et al. \cite{giacintucci06}, hereinafter G06). A fraction 
of $\sim40\%$ is radio quiet at the sensitivity limit of the
1.4 GHz NVSS survey. We inspected all cDs with radio emission 
and selected 13 sources still lacking high sensitivity and 
high resolution radio imaging in the literature. 

For these clusters we performed observations with the GMRT at 1.28 GHz 
to complete the radio information for the whole sample at this frequency.
Furthermore we observed those sources promising for the presence of 
relic emission related to a previous AGN burst at 235 and 610 MHz. 
These low frequencies are suitable for the detection of old 
(i.e. steep spectrum) radio emitting plasma; moreover, combined 
with the 1.4 GHz data, they allow us to study the source spectral 
index (total and point--to--point) over a wide frequency range, providing 
crucial information on the nature of the observed emission.

\section{Preliminary results}

We found 8 active radio galaxies and 5 sources with candidate relic 
emission (steep spectrum and lack of central nuclear activity).
Here we present the GMRT 235 MHz images for three clusters 
and 610 MHz images for two sources of the sample.

\begin{figure}
\centering
\includegraphics[height=2.8cm]{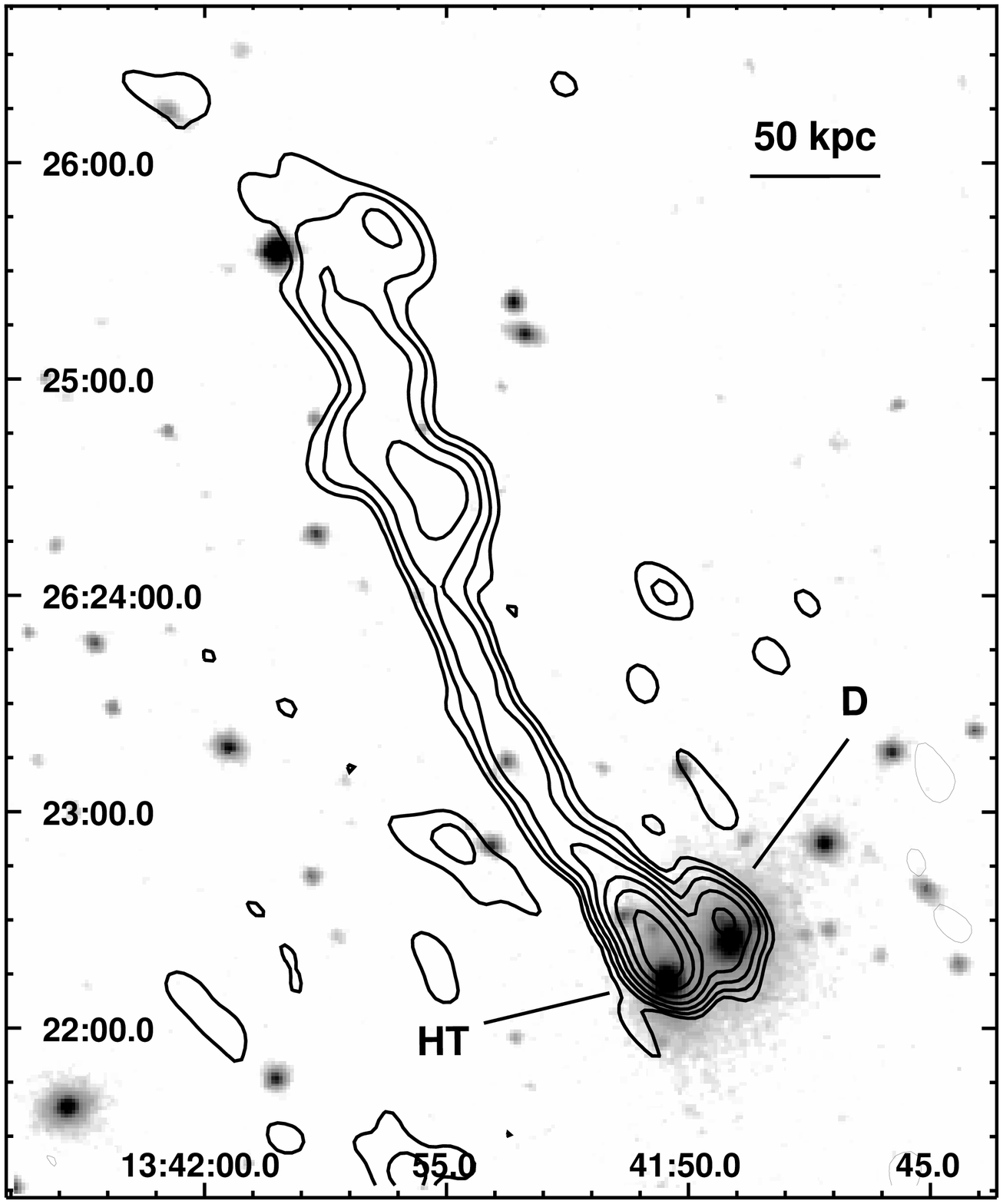}
\includegraphics[height=2.8cm]{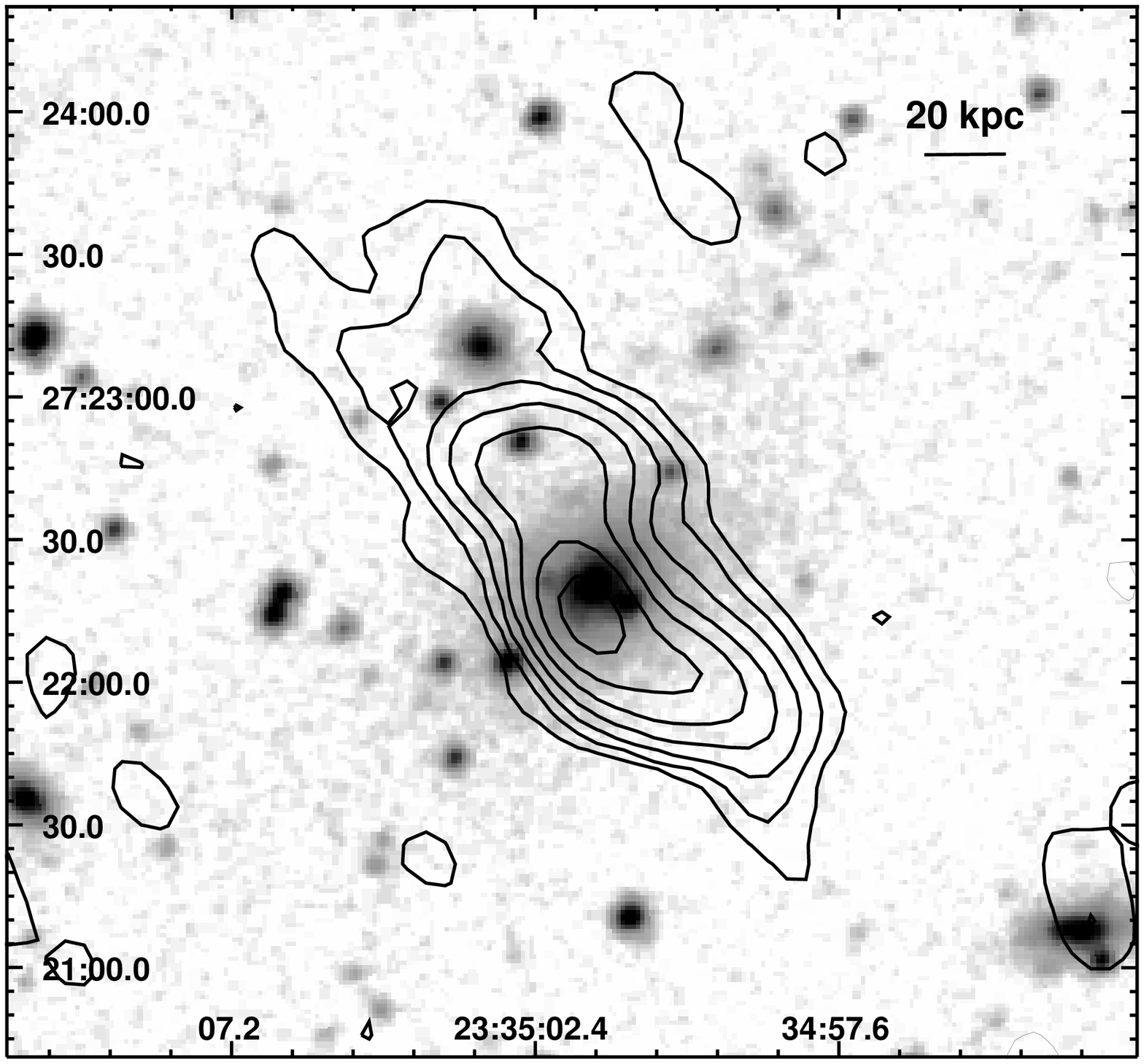}
\includegraphics[height=2.8cm]{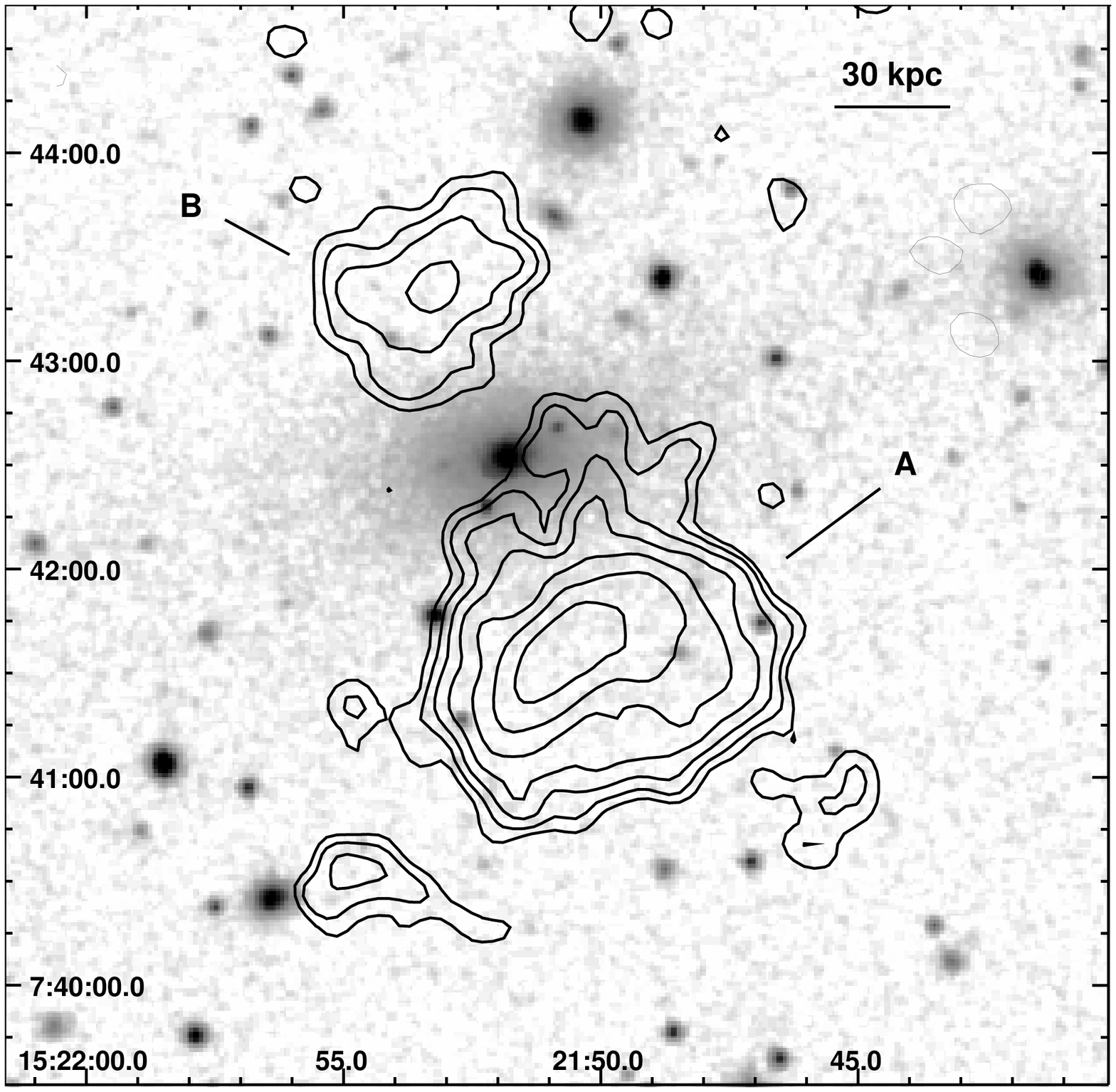}
\caption{GMRT 235 MHz contours (logarithmic, starting 
from $\pm3\sigma$) on the POSS--2 optical image. 
{\it Left panel}: A\,1775, 1$\sigma$=1.5 mJy/b, 
HPWB=12.7$^{\prime \prime} \times 9.1^{\prime\prime}$. 
{\it Central panel}: A\,2622, 1$\sigma$=0.8  mJy/b, 
HPWB=17.1$^{\prime \prime} \times 11.4^{\prime\prime}$; 
{\it Right panel}: MKW\,03s, 1$\sigma$= 2 mJy/b, 
HPWB=12.1$^{\prime \prime} \times 9.1^{\prime\prime}$.}
\label{fig:1}      
\end{figure}

\begin{figure}
\centering
\includegraphics[height=2.8cm]{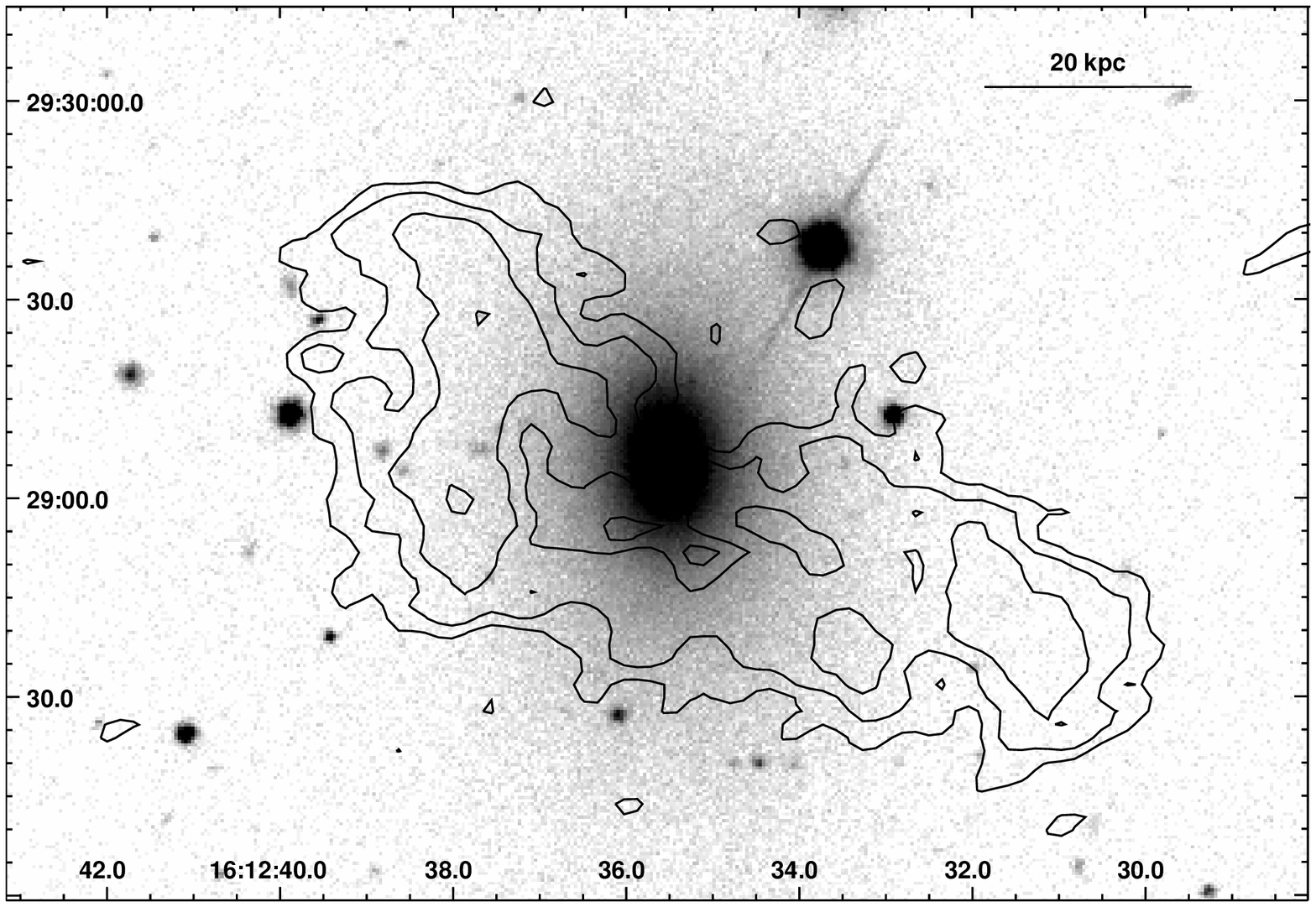}
\includegraphics[height=2.8cm]{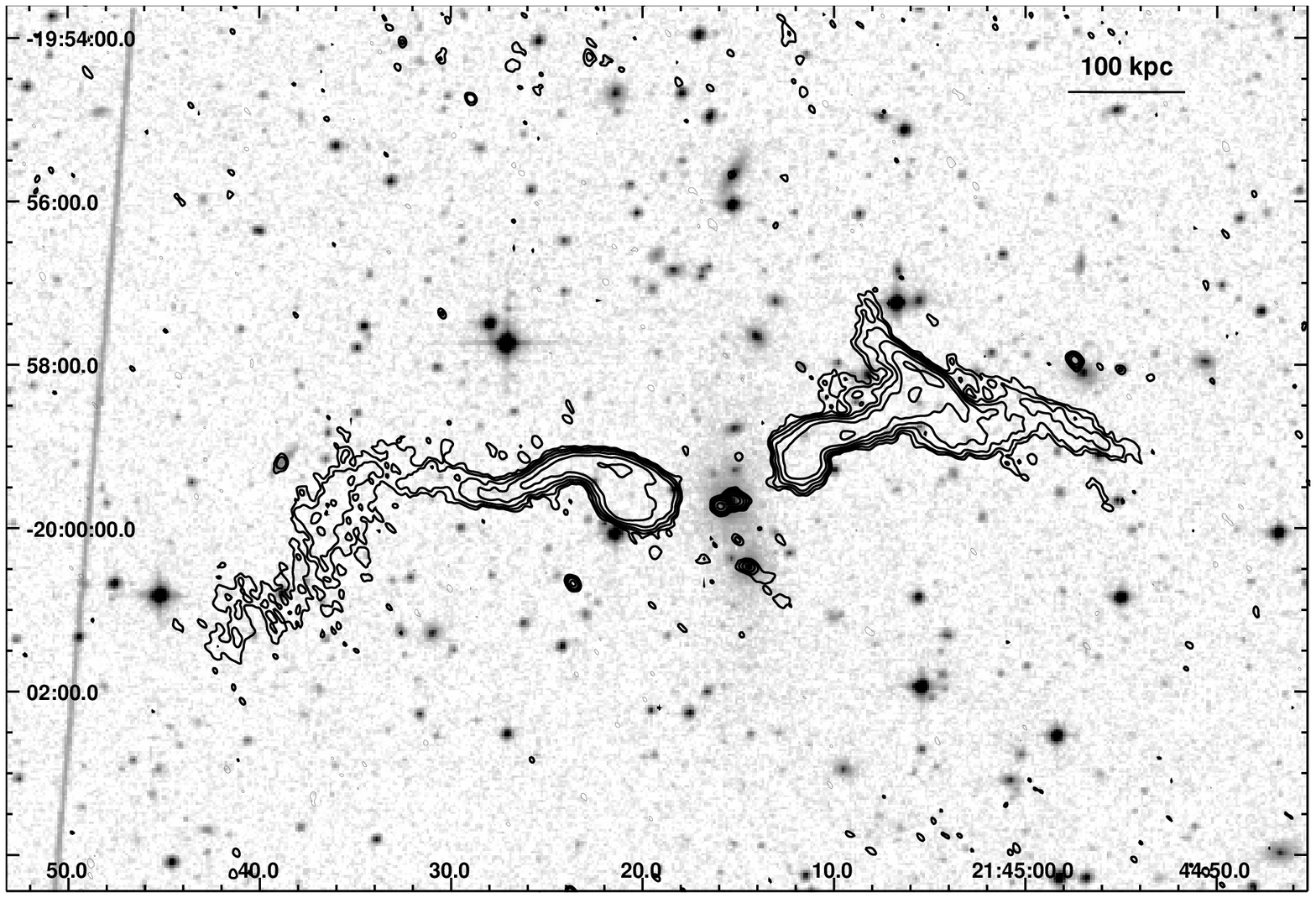}
\caption{GMRT 610 MHz contours (logarithmic, starting 
from $\pm3\sigma$) on the POSS--2 optical image. 
{\it Left panel}: A\,2162, 1$\sigma$=0.15 mJy/b, 
HPWB=7.1$^{\prime \prime} \times 4.9^{\prime\prime}$. 
{\it Right panel}: A\,2372, 1$\sigma$= 0.065 mJy/b, 
HPWB=7.5$^{\prime \prime} \times 5.7^{\prime\prime}$.}
\label{fig:2}       
\end{figure}

\noindent {\bf A\,1775 \index{A1775}} -- A\,1775 hosts a Dumbell galaxy at 
its centre. Both galaxies are radio sources: a 
double source (labelled as D in Fig.\ref{fig:1}) is associated 
with the North--Western component, while an head--tail (HT) 
is associated with the companion galaxy. We measured a total 
flux density of $S_{\rm (HT+D)}$=2.04 Jy at 235 MHz. 
We observed the source also at 610 MHz (G06), and found 
a morphology in very good agreement with the 235 MHz image. 
Using our data, and literature data, we determined a total 
spectral index of $\alpha=1.15$ (S$\propto \nu^{-\alpha}$)
between 80 MHz and 4.85 MHz.
\\
{\bf A\,2622 \index{A2622}} -- A double radio source is associated 
with the cD at 235 MHz (Fig.\ref{fig:1}). 
Our 5$^{\prime\prime}$ resolution observation 
at 610 MHz (G06) reveals a complex structure for 
this source. Its spectral index is very steep ($\alpha=1.42$) 
between 74 MHz and 4.5 GHz. The 235--610 MHz spectral index 
distribution over the source shows that $\alpha\sim0.6-0.7$ in 
the central region, and steepens along the lobes up 
to $\alpha\sim1.8-2$ (G06). This confirms the presence of aged radio 
emission in this galaxy. 
\\
{\bf MKW\,03s \index{MKW03s}} -- A combined GMRT 1.28 GHz/610 MHz 
and {\it Chandra} study of this source was presented in Mazzotta 
et al. \cite{mazzotta04}. 
In the new 235 MHz image (Fig.\ref{fig:2}) the source 
shows two lobes, the southern lobe being brighter than 
the northern one. We do not detect the compact component 
between the lobes observed at higher frequency, and associated 
with the cD nucleus. The source 235 MHz flux density is 
8.44 Jy. The spectral index between 235 MHz and 1.28 GHz 
is very steep ($\alpha=2.49$), confirming the presence of relic 
emission in this source.
\\
{\bf A\,2162 \index{A2162}} -- The 610 MHz radio emission is weak, 
and shows two radio lobes and no hint of a central 
compact component (Fig.2). The source flux 
density at 610 MHz is 240 mJy. Using this value and 
literature data, we found $\alpha=0.81$
in the 74 MHz--4.7 GHz frequency range. 
\\
{\bf A\,2372 \index{A2372}} -- This cD hosts a very large WAT radio 
source, with a total flux density of 1.02 Jy at 610 MHz
(Fig.\ref{fig:2}). We observed A\,2372 also at 235 MHz 
(G06). Using our data and literature values, we found $\alpha$=1.13 
between 235 and 1400 MHz. The image of the 235--610 MHz 
spectral index distribution shows that 
the $\alpha\sim-0.5$ in the core, and steepens 
from $\sim$0.7 to $\sim$1.0 along the jets. In the outer 
regions of the tails $\alpha\sim2$ (G06).

\vspace{-1cm}

\vspace{-1.2cm}
%
%

%
%



\printindex
\end{document}